\documentclass[11pt,aps,amsmath,amssymb,notitlepage]{revtex4-1}
\usepackage{amsmath,amssymb}
\usepackage{slashed}
\usepackage{graphicx}
\usepackage{bm}
\usepackage[top=1.0in,bottom=1.0in,left=1.0in,right=1.0in]{geometry}
\usepackage[colorlinks,linkcolor=blue,citecolor=blue]{hyperref}
\usepackage{subcaption}
\begin{document}

\title{Freeness in cognitive science}

\author{Ewa Gudowska-Nowak}
\email{ewa.gudowska-nowak@uj.edu.pl}
\affiliation{Institute
of Theoretical Physics and Mark Kac Center for Complex Systems Research,
Jagiellonian University, 30-348 Kraków, Poland}

\author{Maciej A. Nowak}
\email{maciej.a.nowak@uj.edu.pl}
\affiliation{Institute
of Theoretical Physics and Mark Kac Center for Complex Systems Research,
Jagiellonian University, 30-348 Kraków, Poland}

\begin{abstract}
In this mini-review, dedicated to the Jubilee of Professor Tadeusz Marek, we highlight in a popular way  the power of so-called free random variables (hereafter FRV) calculus, viewed as a potential probability calculus for the XXI century, in applications to the broad area of cognitive sciences. We provide three examples: (i) inference of noisy signals  from multivariate correlation data from the brain; (ii) distinguished role of non-normality in real neuronal models; (iii) applications to the field of  deep learning in artificial neural networks. 

\end{abstract}
\maketitle
\newcommand{\dg}{\dagger}
\newcommand{\btr}{\textup{bTr}}
\newcommand{\<}{\left<}
\renewcommand{\>}{\right>}
\newcommand{\idm}{\mathbf{1}}
\newcommand{\zb}{\bar{z}}
\newcommand{\wb}{\bar{w}}
\newcommand{\vb}{\bar{v}}
\newcommand{\gb}{\bar{g}}
\newcommand{\linia}{\rule{\linewidth}{0.4mm}}
\newcommand{\unit}{\mathbb{1}}
\newcommand{\cG}{{\cal G}}
\newcommand{\cQ}{{\cal Q}}
\newcommand{\cX}{{\cal X}}
\newcommand{\cY}{{\cal Y}}
\newcommand{\cA}{{\cal A}}
\newcommand{\cR}{{\cal R}}
\newcommand{\cB}{{\cal B}}
\newcommand{\cW}{{\cal W}}
\newcommand{\cL}{{\cal L}}
\newcommand{\braket}[2]{\< #1 | #2 \>}
\newcommand{\bb}[1]{{\bf #1}}
\newcommand{\re}{\textup{Re}}
\newcommand{\im}{\textup{Im}}
\newcommand{\be}{\begin{eqnarray}}
\newcommand{\ee}{\end{eqnarray}}
\newcommand{\MeijerG}[8][\bigg]{G^{{ #2 },{ #3 }}_{{ #4 },{ #5 }} #1( \begin{matrix} #6 \\ #7 \end{matrix}\, #1\vert\, #8 #1)}

\section{Introduction}
Cognitive science is a broad domain of  interdisciplinary research, dedicated to the ultimate understanding of  mind and intelligence. Within the last hundred years, it has passed a dramatic change from mostly  humanistic (and even philosophical) area to the domain of hard science.  This transition was caused, among others,  by several  technological breakthroughs in deciphering {\it in vivo} the neuronal signals~\cite{FAFROWICZ}. Techniques alike dense array encephalography (dEEG) 
(including more invasive version alike electrocorticography), functional magnetic resonance imaging (fMRI), (including diffusive tensor analysis), magnetoencephalography (MEG), transcranial magnetic stimulation (TMS), and finally, optogenetics,  provided gargantuan amount of data at wide  spectrum of temporal and/or spatial resolutions. That flood of data has  opened the door for  the methodologies of empirical sciences, especially in the area of complex  systems. Standard mathematical  tools in this area, due to the noisy character of the data,  include probability calculus and stochastic differential equations. However, the new challenge with respect to modern neurological data comes from their high dimensionality - e.g. number of voxels in fMRI  single snapshot  reaches tens of thousands and   time series length for dEEG recordings can be easily order of magnitude larger, since one can probe up to thousand signals per second. This multivariate character of time series  brings immediately the connotation to random matrix theory. Already, more then 90 years ago John Wishart~\cite{WISHART} has asked  the following question: It is well known, that the sum of squares of the independent Gaussian  variables $x_t$, where $t$ labels e.g. times of the measurement $(t=1,...,T)$,  is given by $\chi^2$ distribution. What is the generalisation of $\chi^2$ distribution, if  we look at similar process, but with multivariate Gaussian variable $x_{it}$, where indices $i$ count the number of different measured objects?  For example, in our  setup, $i$ may correspond to distinct electrodes on the scalp ($i=1,...,N$). The resulting distribution is today known as Wishart distribution, corresponding to the distribution of {\it correlation matrix}    $C_{ij}=\frac{1}{T} \sum_{t=1}^T x_{it}x_{jt}=\frac{1}{T} {\rm tr} XX^{\tau}$, where in the last equality we have exploited the matricial notation -  here $x_{it}$ is the element of $N \times T$ matrix $X$ and $\tau$ denotes the transposition of such  matrix.  One can simply say, that random matrix theory 
is just the sort of probability theory, where random variable is matrix-valued.  In the 50-ties of the previous century, random matrix theory started making the impact on almost all  branches of physics, and later, on other domains of hard science. The reason was that people like Wigner,  Porter, Dyson, Gaudin and Mehta started looking at the statistics of eigenvalues of random matrices   instead of the statistics of the elements of matrices~\cite{TAO}. It turned out, that at the microscopic level (spacing between eigenvalues scaling like $1/N$, where $N$ is the dimension of the matrix), the spectral properties are universal and,   in general, independent  on the probability distribution of the matricial elements. On the other side, at the macroscopic level, the resulting spectral laws started to tend to simple expressions in the limit when the size of the matrix was very large. 
In 90-ties of the previous century, this observation was formalised in mathematical language by Voiculescu~\cite{VOICULESCU}, leading to the emergence of free random calculus, perhaps the most fundamental and versatile generalisation of probability theory for non-commuting random operators. Since Voiculescu theory  agrees with RMT   in the limit of infinite dimension of random matrices, it represents an asymptotic limit. But, first, the convergence to the asymptotic results is very fast, even for moderately large matrices of dimensions of order 10 ( this is sometimes anecdotally expressed as $8 \approx \infty$), second, analysed matrices have dimensions easily reaching $10^3$, so deviations from asymptotic results are usually very small, therefore  the method of FRV is operational and practical.   This is why, from the probabilistic point of view, we do not hesitate to call   FRV calculus as the probability calculus for the XXI century. 

\section{Free Random Variables in a Nut-Shell}
Free random variable calculus  can be viewed as a generalisation
of classical probability calculus, for the case of non-commuting operators (viewed here as infinite, noncommuting  matrices), hence it is natural to explain the cornerstones of  FRV using the intuition from classical probability calculus. 
Let is consider the problem of "adding" two, {\it independent} random variables $x_1$ and $x_2$, from corresponding distributions $p_1(x_1)$ and $p_2(x_2)$, i.e. the problem of finding distribution
\begin{equation}
 p(s)=\int dx_1dx_2p_1(x_1)p_2(x_2)\delta (s-(x_1+x_2))=\int dx p_1(x)p_2(s-x)   
\end{equation}
The last equation represents the convolution, so it is natural to take Fourier transform of the probability distribution $p(s)$. 
Then $\hat{p}(k)=\int p(s)e^{iks}ds= \hat{p}_1(k)\cdot \hat{p}_2(k)$. 
Expanding the exponent we see, that Fourier transform generates moments of the distributions - $\hat{p}(k)=\sum_{n=0}^{\infty} \frac{(ik)^n}{n!} m_n$ with $m_n=\int dx x^n p(x)$. Further simplification happens when we take the logarithm of  Fourier transform, $r(k)=\ln \hat{p} (k)$. 
Then we have an addition law 
\begin{equation}
    r_{1+2}(k)= r_1(k) +r_2(k)
\end{equation}
Since $r(k)$  can be also viewed as another generating function
$r(k)=\sum_{n=1}^{\infty}\frac{(ik)^n}{n!} \kappa_n$, where coefficients $\kappa_i$ are called cumulants, we just have found the way of linearizing convolution of independent distributions: first, we calculate the cumulants of individual components, then we add them algebraically. Resulting series yields the cumulants of the convolution.  Particularly simple example is provided by the Gaussian
 $p(x)=\frac{1}{\sqrt{2 \pi}}e^{-x^2/2} \equiv N(0,\sigma^2=1)$. Fourier transform is also a Gaussian, $\hat{p}(k)= e^{-k^2/2}$, and the $r(k)=-k^2/2$. We see that all cumulants vanish except of the second one, $\kappa_2=1$. 
The convolution of two standard Gaussians is therefore also a Gaussian, but with a dispersion $\sigma^2=1+1$.

Now we will parallel the above reasoning in the case of infinitely large random matrices. We start from the symmetric random matrices,  since their spectrum is real.  We define first  the analogue of independence, which is called freeness.  Consider two large diagonal matrices of size $N$ by $N$, named $X$ and $Y$. There are not free. However, if at least one of them we rotate by Haar measure, $Y \rightarrow OYO^{\tau}$, where $O$  is random orthogonal transformation, then matrices $X$ and $OYO^{\tau}$ are mutually free in the limit of infinite size of matrices. Intuitively,  freeness is equivalent to maximal decorrelation of corresponding sets of eigenvectors.  As a next cornerstone we introduce the moment generating function (Green's function), defined as 
\begin{equation}
   G_X(z)=\int \frac{\rho_x(\lambda)}{z-\lambda} 
\end{equation}
where $\rho_x(\lambda) \equiv \lim_{N \rightarrow \infty} \frac{1}{N} <\sum_i \delta(\lambda-\lambda_i > $ is the average spectral density of the matrix $X$ with respect to the probabilistic measure $<...>=\int DX ...e^{-N {\rm tr} V(X)} $ therefore an analog of the probability density function $p(x)$ in classical probability. For example, Gaussian measure corresponds to $V(X)=\frac{1}{2}X^2$. Note, that for large complex values $z$, $G_X(z)=\sum_{k=0}^{\infty}z^{-k-1} m_k$, where spectral moments read $m_k=\int \lambda^k \rho_x(\lambda) d\lambda= \frac{1}{N} <{\rm tr} X^k>$. 
Finally, we define the function generating the free cumulants. 
In FRV calculus this function is called R-transform and is defined as $R(z)=\sum_{k=1}^{\infty} \kappa_k z^{k+1}$.  Its relation to $G$ is involved - basically, modulo the shift $1/z$, it is the functional inverse (for any $z$) of the Green's function - 
$G[R(z)+\frac{1}{z}]=z$ (or equivalently $R[G(w)]+\frac{1}{G(w)}=w$).
The algorithm of "addition" of the spectra  is now as follows.\\
(i) Knowing $\rho_X(\lambda)$ and  $\rho_Y(\lambda)$, we construct the corresponding Green's functions $G_X(z)$ and $G_Y(z)$\\
(ii) We invert functionally both Green's functions, finding $R_X(z)$ and $R_Y(z)$ \\
(iii) We perform the addition law $R_{X+Y}(z)=R_X(z) +R_Y(z)$, and we functionally invert the result, getting $G_{X+Y}(z)$.\\
(iv) Finally, we reconstruct $\rho_{X+Y}(\lambda)$, using the analytical properties of $G_{X+Y}(z)$
\begin{eqnarray}
-\frac{1}\pi {\rm Im }  \lim_{\epsilon \rightarrow 0}  G_{X+Y}(z)|_{z=\lambda +i\epsilon}&=& \lim_{\epsilon \rightarrow 0}
\int \rho_{X+Y}(\mu) \frac{1}{\pi} \frac{\epsilon}{(\mu-\lambda)^2 + \epsilon^2}d\mu \nonumber \\ 
&=& \int \rho_{X+Y}(\mu)\delta(\mu-\lambda)d\mu =\rho_{X+Y}(\lambda)
\end{eqnarray}
We conclude this part by providing a pedagogical ensemble. Let us consider "the Gaussian" in the FRV calculus. By analogy to  classical case, we consider the case, when only one cumulant is non-vanishing, i.e. $\kappa_2$, which for simplicity we put to 1. Then $R(z)=z$, and inverting the R-transform reduces to the solution of quadratic equation $G+1/G=z$. The solution with proper asymptotic behaviour for large $z$ reads $G(z)=\frac{1}{2}(z-\sqrt{z^2-4})$, and rerun of the argument (iv) from above list yields $\rho(\lambda)=\frac{1}{2\pi}\sqrt{4-\lambda^2}$, i.e. the famous Wigner semicircle. "Addition" of mutually free semicircles parallels the "addition" of independent Gaussians  in classical probability.  

We conclude this introduction with few comments on multiplication of random variables. In classical probability, at least formally, multiplication is not very much different from addition, due to the relation $e^x \cdot e^y=e^{x+y}$. If $x,y$ would be replaced by large matrices $X,Y$, above relation does not hold, since matrices, in general do not commute. Even worst, the product of two symmetric matrices is usually not symmetric, which means, that first, the spectrum is becoming complex-valued, second, the eigenvectors do not decouple from the spectrum and are crucial for analysing e.g. the stability problems. Luckily, there exist few cases, where the application of the whole machinery for non-normal random matrices is not necessary.  First, consider  the case, when we multiply two matrices, where at least one of them is positive.  One can then define so-called S-transform, which is multiplicative (i.e. $S_{X\cdot Y}(z)=S_X(z)\cdot S_Y(z)$ and S-transform is  related to R-transform by
$S_X(z)R_X(zS_{X}(z))=1$, which allows to extend the addition program for multiplication. 
Second exception  corresponds to the case, when random matrix $X$ can be decomposed as $X=PO$ where $P$ is positive, $O$ is a Haar measure and both $P, O$ are mutually free. In such case, the spectrum  has azimuthal symmetry, and only radial distribution is
non-trivial $\rho(r)=\frac{1}{2\pi r} dF(r)/dr$, ( where $r=|\lambda|$ and $F(r)$ is the cumulative radial distribution), so this case corresponds to a quasi-one dimensional case. In this case, powerful Haagerup-Larsen (or single ring) theorem holds~\cite{HL} for the spectra, 
\begin{eqnarray}
 S_X(F(r)-1)&=&\frac{1}{r^2} \nonumber \\
 O_X(r) &\equiv& \frac{1}{N^2} \left<\sum_i \delta^{(2)}(z-\lambda_i) <L_i|L_i><R_i|R_i>\right>= \frac{1}{\pi r^2} F(r)(1-F(r))
\end{eqnarray}
whereas second line  addresses the eigenvectors~\cite{NS}, i.e.  $<L_i|X=<L_i| \lambda_i$ and $X|R_i>=\lambda_i|R_i>$  are left and right, distinct eigenvectors corresponding to same complex-valued eigenvalue $\lambda_i$.  Last but not least, if the problem of finding eigenvalues and eigenvectors of non-normal operator $X$ is very hard, one may look at the singular value decomposition (SVD), i.e. consider the real spectrum of the operator $XX^{\tau}$. In below, we will exploit all three above mentioned  special cases corresponding to multiplication laws for large random matrices. 

\section{Spectral analysis of correlation matrices}
We explain the main idea of spectral analysis of correlation matrices on the basis of anecdotal example. Let us consider the measurement of dense array electroencephalogram performed on one of the authors of this review, in so-called resting state {\it dolce far niente} (Figure~1).  We specify the number of electrodes $N$ and the elapsed time of the measurement. Since the measurement is done at fixed intervals (e.g. with  frequency 100Hz), as a result we obtain the multivariate time series $M_{it}$ of measurements of $i=1,...,N$ electrodes at $T$ time steps $t=1,...,T$, with $T$ much larger than $N$. Let us now look at the fluctuations between the consecutive measurements $(X_{it}=M_{it}-M_{i,t-1}$), and let us   standartize these fluctuations, i.e. for each electrode we calculate  the mean and the variance of the time series $X_{it}$, and  for each number of electrodes we calculate 
$x_{it}= (X_{it}-<X_i>)/\sqrt{<X_i^2>}$. Finally, we  construct the Pearson estimator for the correlation matrix 
\begin{equation}
    C_{ij}=\frac{1}{T} \sum_{t=1}^T x_{it}x_{jt}
\end{equation}
In matricial form, above equation reads $C=\frac{1}{T}XX^{\tau}$. In Figure~2 we plot  the histogram of all eigenvalues of this estimator (in orange).  The power of FRV calculus stems from the fact, that we can easily construct analytical benchmarks, which allow the comparison of the measured data with some assumptions on the nature of true correlations, and then, perform the inference of the true correlation from the data.  Let us  start from the simplest assumption, that all $x_{it}$  come from independent  central, standard, Gaussian distributions $N(0,1)$.  The correlation estimator  in this case is just the Wishart matrix $C_W$. In the case when $N=T>>1$, the simple inspection shows that all spectral cumulants $\kappa_i$ are identical and equal to 1. We can therefore consider the  resulting spectral distribution as an analogue of Poisson distribution in classical probability. 
The resulting R transform  is therefore $R(z)=\sum_{i=0}^{\infty}\kappa_i z^i=\sum_{i=0}^{\infty} z^i=\frac{1}{1-z}$. When $T>N>>1$, the cumulants are  simply rescaled by the "rectangularity" $r=N/T$, leading to  $\kappa_i=r^i$, so the R-transform for Wishart reads  
\begin{equation}
  R_W(z)=\frac{1}{1-rz}  
  \label{Wishart}
\end{equation}
Since, by definition $R(G(z))+1/G(z)=z$, formula (\ref{Wishart}) leads to a quadratic algebraic equation for the Green's function, with obvious solution for $G_W(z)$.  Taking the imaginary part of the solution leads to spectral density for the Wishart ensemble
\begin{equation}
\rho_{MP}(\lambda)=\frac{1}{2\pi r \lambda} \sqrt{(\lambda_+-\lambda)(\lambda-\lambda_-)}
\label{MP}
\end{equation}
where $\lambda_{\pm}=(1\pm \sqrt{r})^2$. This is the celebrated Marcenko-Pastur distribution.  This famous formula is a benchmark of lack of any correlations in the measured multivariate time series.  Why then this spectrum does not correspond to the spectrum of true covariance matrix for multivariate Gaussian (which is unit diagonal), or, in other words, to the spectral measure $\rho_{true}=\delta(\lambda-1)$? The reason is the finite number of measurements, which always introduces the noise. Note, that only in the limit $T\rightarrow \infty$ with $N$ fixed, the Marcenko-Pastur distribution tends to single eigenvalue 1, since 
the support of the spectrum, $[\lambda_-, \lambda_+]$ shrinks in the limit $r \rightarrow 0$ to this value from both sides of the support.  
Still, the disagreement between orange  histogram  and analytical result  (blue line) from Marcenko-Pastur distribution shows that even in the resting state the electric activity of the brain of the author is more involved comparing to Gaussian noise, which is reassuring!  Finally, left us make the next "measurement".  Let us now destroy all the temporal (causal) correlations in the measured EEG data of the author, by multiple reshuffling of all the columns  in matrix $X$.  Then, we construct again the covariance matrix and calculate the spectrum. As expected, the data now  (blue histogram on Figure~2) are in perfect agreement with pure noise data, i.e. with the Marcenko-Pastur distribution,  since all causal correlations have been  destroyed.  
\begin{figure}
\centering
\begin{minipage}{.485\textwidth}
  \centering
  \includegraphics[width=.38\linewidth]{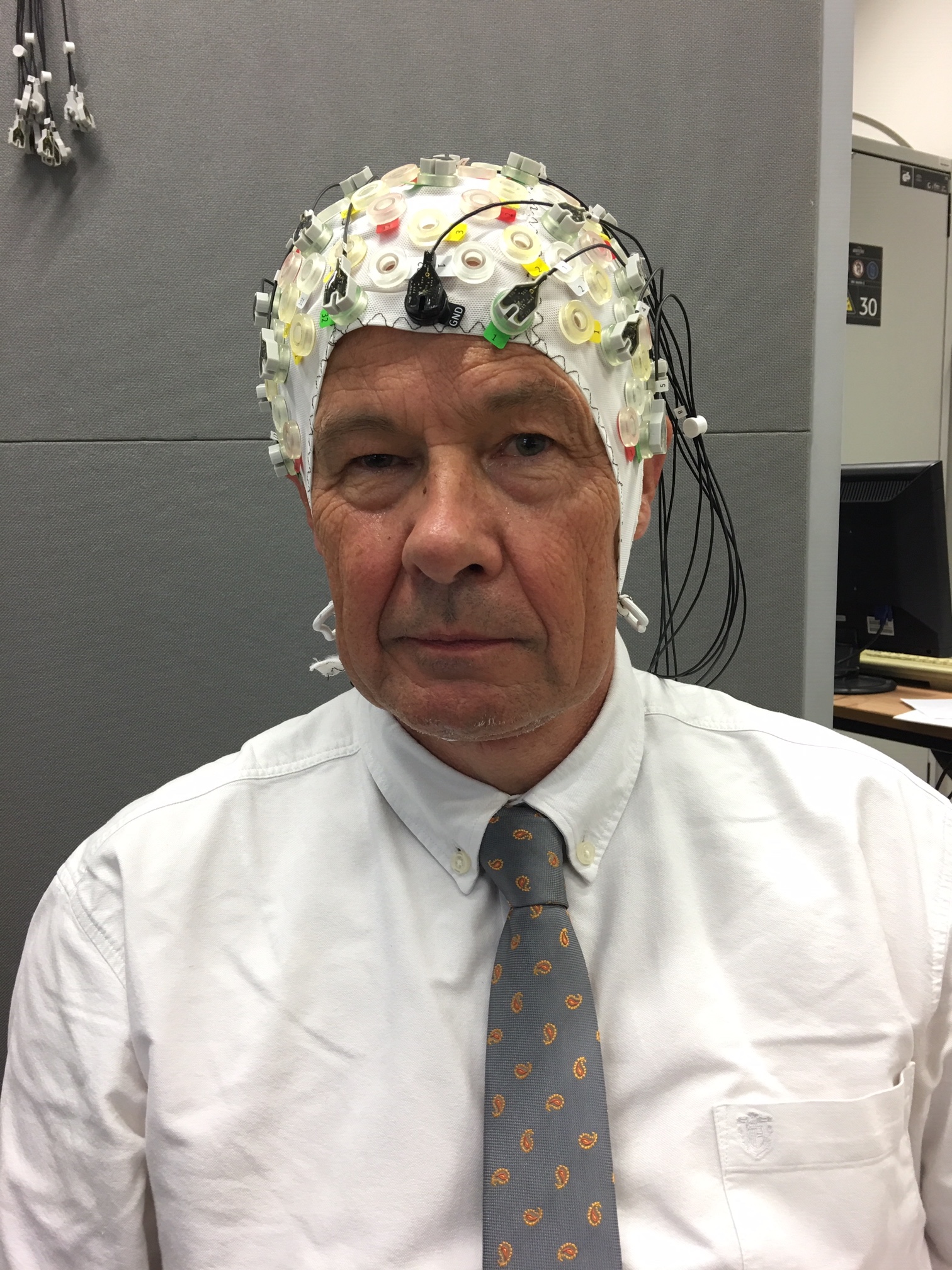}
  \caption{dAEEG experiment, done at J. Ochab's lab.}
  \label{fig:test1}
\end{minipage}%
\begin{minipage}{.49\textwidth}
  \centering
  \includegraphics[width=.68\linewidth]{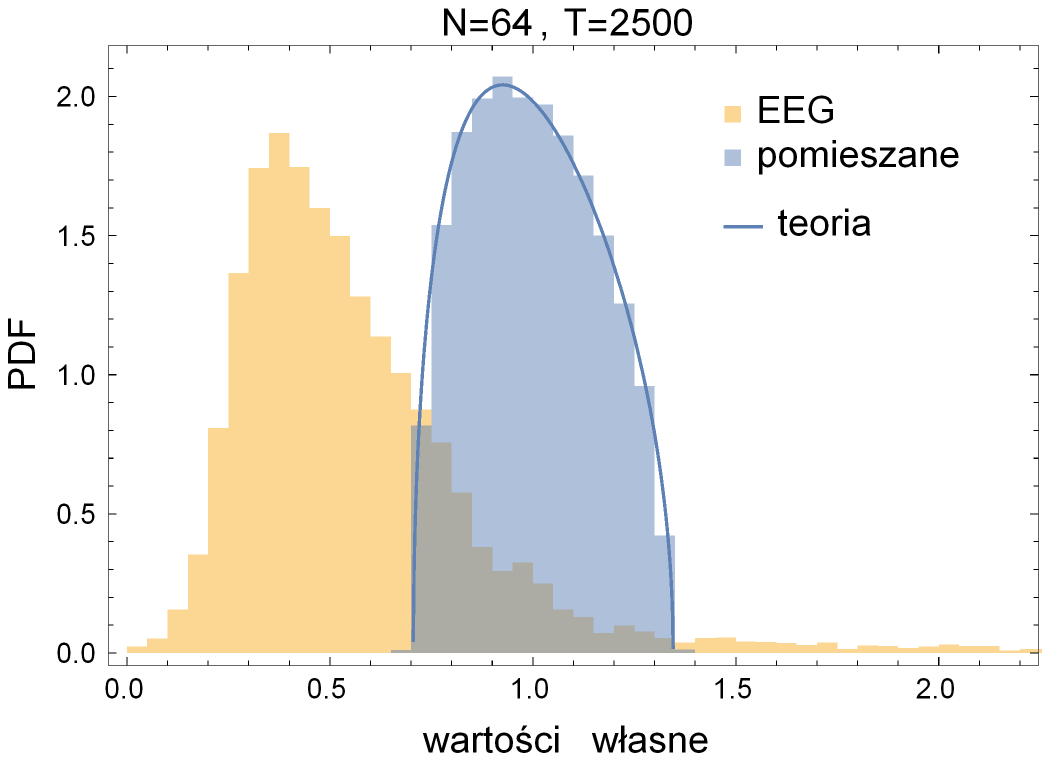}
  \caption{Spectral histograms of the pdf of Pearson  estimator.}
  \label{fig:test2}
\end{minipage}
\end{figure}

Of course, inferring  the information that the signals in the head of one of the authors are not pure noise is not very far reaching. 
In general, we expect that the  covariance matrix is much more sophisticated, e.g. $ < x_{it}x_{js}>=A_{ij}B_{ts}$, where matrices $A$  reflect the true correlations between the clusters of electrodes and matrices $B$ reflect temporal (auto)correlations for the same electrode. Assuming still the Gaussian character of the fluctuations,  we are facing the problem of calculating spectral moments (cumulants) of the type $<{\rm tr} [XX^{\tau}]^k>_{cW}$, where index $cW$ means correlated Wishart, i.e. the true measure is proportional to 
$\exp {-\frac{1}{2} {\rm tr} (A^{-1}XB^{-1}X^{\tau})}$. Now  we can see  the power of FRV calculus. Let us change the variables $\sqrt{A^{-1}}X\sqrt{B^{-1}} \equiv Y$. Note that this change of variables converts the measure into pure Gaussian one,  $\exp {-\frac{1}{2} {\rm tr} Y Y^{\tau}}$, but at the expense of complicating the moments, which  read now  $<{\rm tr} [\sqrt{A}YBY^{\tau}\sqrt{A}]^k>_{W}= <{\rm tr} [AYBY^{\tau}]^k>_{W}$, where we used the cyclic property of the trace. Such mixed moments  can be easily disentangled using the powerful S-transform technique. First, one can "factorise"  the spectrum of $A$ from the moments  $<{\rm tr} [YBY^{\tau}]^k>_{W} \sim <{\rm tr} [BY^{\tau}Y]^k>_{W}$, where again we used the cyclic properties of the trace. Second, one can "factorise" further  the spectrum of $B$ from anti-Wishart  moments $<{\rm tr} [Y^{\tau}Y]^k>_{W}$, which are equal, modulo trivial normalisation to moments of Wishart 
$<{\rm tr} [YY^{\tau}]^k>_{W}>$. In such way we have spectrally disentangled the correlated moments in terms of elementary spectral constituents. The resulting explicit formulae are complicated (so we do not list them),  but there are exact, and allow to infer the true moments from the measured estimators. Assuming  the a priori unknown structure of correlations $A$ and/or $B$ and minimising the error  allows  the explicit optimisation of the predictions for the true correlation matrices. For explicit ensembles, we refer to~\cite{QF}. 
The additional advantage of the FRV spectral methods stems from the fact, that they  can be easily generalised for other cases of randomness (L\'{e}vy, Student-Fisher etc) and can be applied also in the case of non-linear estimators, alike nonlinear shrinkage estimators.

 \section{Rajan-Abbott model for real neuronal network}
 In the majority of  models of synaptic interactions,  strength of interactions between all the pairs of $N$ neurons is provided by random adjacency matrix. The neuroscience imposes however stronger constraints  comparing to random matrix theory, in particular,  allows at least two types of neurons to be active, excitatory and inhibitory ones, with two different values of means of variances from e.g. the simplest Gaussian ensembles. The synaptic matrix can be therefore mimicked as $X=M+W$, where $W=G\Lambda$, with $G$ purely random (Gaussian Ginibre ensemble type), and $\Lambda$  the diagonal  with first elements  corresponding to excitatory neurons equal to variance $\sigma_E$ and remaining inhibitory neurons with variance $\sigma_I$. One rank matrix $M$ includes the information on the means $\mu$ of two kinds of neurons. Since empirical studies show that the amount of inhibition and excitation of a neuron is the same even at the scale of few milliseconds, global constraint is applied, 
$f_E\mu_E+f_I\mu_I=0$, where corresponding $f_i$ are the fractions of pertinent neurons. Even in this case, the non-normal character of the spectrum causes the eigenvalues of $X$ to be dramatically different from simple assumption of Gaussianity of  $W$. In their important contribution Rajan and Abbott~\cite{RA} suggested an additional local condition,  demanding that the sums of 
strengths coupled independently for each neuron vanish.  In recent paper~\cite{EWARJ}, we have reanalysed that analysis using the powerful tools of FRV calculus, exploiting the Haagerup-Larsen theorem and the fact, that if $G$ is R-diagonal, the product of $G\Lambda$ is as well. 
First, we have provided back-on-envelope re-derivation of the original model, using the advantage  of FRV variables. Second, we have addressed the issue of the statistics of  eigenvectors, which was not amenable in original formulation. Main message was, that left-right eigenvector correlation $O_X$  (eq. 5) is dramatically sensitive to the local balance condition. Since FRV calculus allows also the calculations where moments do not exist, we considered this case, showing that in the case of heavy-tailed spectra the above effect is magnified by orders of magnitude.  
This means, that the full description  of dynamical processes of realistic adjacency neuronal networks requires the entangled dynamics of both eigenvalues and eigenvectors, contrary to the evolution of normal (symmetric matrices), where eigenvectors decouple. For balanced networks,  the sensitivity of eigenvalues to any additive perturbation  is dramatic, which calls for some specific, powerful  mechanism for the stabilisation of the spectra of adjacency matrices.  We have envisioned,  that the generic mechanism of such type  can be provided by the transient behaviour~\cite{GRELA}. We notice, that such mechanism is consistent with the model of del Molino et al.~\cite{MOLINO}  et al. 

\section{Freeness in Deep Learning}
Free random variables applications to deep neural networks was pioneered by Google AI team~\cite{PENN},  where the particular, generic  fit to tailor the initialisation in feed forward networks was obtained (so-called isometry).  In this section, we briefly advertise the extension of above construction to the case of residual networks~\cite{OURRESNET}.
In residual network, the information propagates  according to the prescription
\begin{eqnarray}
{\bf x}^l&=&\phi({\bf h}^l) +a {\bf h}^{l-1} \nonumber \\
{\bf h}^l&=& {\bf W^l} {\bf x}^{l-1} +{\bf b}^l
\end{eqnarray}
where $l$ runs the depth of the network ($l=1,,,,,L$),  and $N$-dimensional  (here $N$ - number of neurons in each layer) vectors ${\bf h, x}$  are pre- and post-  activations  for each layer.  Here, for fixed layer, ${\bf W}$ is the synaptic matrix, $\phi$ is a generic, non-linear, activation function and ${\bf b}$ are real valued bias vectors. Parameter $a$ tracks the influence of skip connections in the networks. In the process of adjusting the weights during training, the crucial role is played by the Jacobian of transition from one layer to the next one, i.e. 
\begin{eqnarray}
\frac{\partial x_k^{l}}{\partial x_t^{l-1}}=\left[{\bf D}^{l}{\bf W}^l +a{\bf 1}  \right]_{kt}
\end{eqnarray}
where ${\bf D}^l$ is a diagonal matrix  $D_{ij}^l=\phi^{'}(h^l_i)\delta_{ij}$. Note that too large or too small gradients in the Jacobian matrix will harm the learning process, leading either to chaos or to un-effective learning, respectively.  
The total input-output Jacobian is the product of Jacobians for each $L$ layers of the network, and  has the form of the matrix 
\begin{eqnarray}
{\bf J}=\prod_{l=1}^L ({\bf D}^l {\bf W}^l +a{\bf 1})
\end{eqnarray}
Understanding the spectral properties of such object is of paramount importance. Luckily, one can address this problem using FRV, since the initialisation of the weight matrix ${\bf W}$ is usually Gaussian. The general structure resembles the non-hermitian multiplicative diffusion~\cite{GJJN}. Since in this case the spectrum is complex, it is technically easier to consider SVD, i.e. to study the real spectrum of ${\bf J} {\bf J}^{\tau}$. 
Let us start from the simplest example, when we put $a=0$ and ${\bf D}={\bf 1}$. This is a linear problem of understanding the spectral properties of the SVD of the product ${\bf W}_i...{\bf W}_L$ of random matrices. Luckily, we can use the power of FRV calculus, noticing that the spectral properties of the product of such matrices is equivalent to the spectral properties of the $L$-power of the single random matrix ${\bf W}$~\cite{NowakBurdaSwiech}.  Then, the simple application of Haagerup-Larsen theorem  shows, that one can just change the variables 
\be
S_{{\bf J}^{\tau}{\bf J}}(F_{\bf J}(r)-1)=\frac{1}{r^{2/L}}
\ee and the   crucial parameter  $\xi=\frac{1}{N} <{\rm tr} {\bf W}^{\tau} {\bf W}>$ is just the outer rim of the famous single ring theorem. 
In the case of non-linear case, the similar reasoning holds, following our argument from the previous section, that the product of R-diagonal ${\bf{W}}$ and any other matrix ${\bf D}$ is still R-diagonal.  
 So,  we can use again Haagerup-Larsen theorem and simple change of variables.  This is the mathematical essence of  Google AI team observation. 
The crucial object is, as before, the outer rim of the single rim theorem, which reads now 
\begin{eqnarray}
\chi=\frac{1}{N} <{\rm tr} ({\bf DW})^{\tau}{\bf DW}>
\end{eqnarray}
The main  observation of the Google AI team was, that even at the outer rim of single rings theorem, the  value of the maximal SVD eigenvalue still grows with the depth of the neural networks $L$. Therefore for e.g. both ReLU  and hard-tanh networks,  there is no way that any choice of Gaussian initialisation  can prevent  the failure  of dynamical  learning procedure. However, in the case when initialisation was  based on Gaussian {\it orthogonal} random  matrices, similar rerun of arguments  has shown, that e.g.  for hard-tanh  networks such fine tuning was possible, even for very large $L$. This spectacular agreement of numerical simulations compared to  theoretical predictions based on FRV calculus was the first demonstration of the power of FRV techniques in Machine Learning. 

The generalisation for ResNet networks ($a\neq0$) is non-trivial, since the shift in multiplication process destroys the azimuthal  symmetry of the spectrum, and invalidates the assumptions of  the Haagerup-Larsen theorem. Nevertheless, more sophisticated  tools of FRV calculus still can be used~\cite{OURRESNET}, leading to the isometry also in the case of ResNets. In particular, for several different activations functions (e.g. ReLU, tanh, hard tanh, sigmoid,  SeLU, leaky ReLu), proper rescaling  of  initialisation conditions lead to isometry (scaling). Analytical results based on FRV calculus were confronted with numerics based on CIFAR10 datasets, confirming the power of FRV when  applied to Deep Learning.

\section{Summary}

In this mini-review, we have highlighted three different aspects 
of cognitive data analysis using the modern tools of FRV calculus.  Our motivation was two-fold. First, we wanted to stress, how broad is the spectrum of FRV tools when applied to different neuroscience datasets. The second motivation is however deeper. 
Nowadays, the areas of statistical analysis of human (or mammalian, to be more general) Big Data  brain  networks,  simulations of real, often low-level  neuronal systems alike considered here Rajan-Abbott model  and  an exploding area  of artificial neural networks (deep learning in ML) have little in common, despite  obvious general motivations to better understanding  how to   emulate (outperform?)  the human mind. On top of semantic differences, all three areas are using different tools and different mathematical formalisms, sometimes at very different level of mathematical rigidity.  In our opinion, FRV calculus provides a rare opportunity for {\it scientifically more rigid} comparison  of these three so different aspects of understanding  the broad empirical spectrum of cognitive sciences, at the level  when verification of hypotheses  and prospects of assessing new algorithms  based on bio-inspiration  can be verified at the  quantitative level.

%
%


%
%

%
%

\section*{Acknowledgments}
 The research was supported   by the TEAMNET POIR.04.04.00- 00-14DE/18-00 grant {\it "Bio-inspired Artificial Neural Networks"} of the Foundation for Polish Science and by the Priority Research Area Digiworld under the program Excellence Initiative -- Research University at the Jagiellonian University in Krak\'{o}w.

\end{document}